\documentclass[twocolumn, 11pt]{article}
\usepackage{array, float}
\usepackage[left=18mm, right=18mm, top=26mm, bottom=24mm]{geometry}
\usepackage{newtxtext}
\usepackage{amsmath, amssymb}
\usepackage{mathrsfs}
\usepackage{tabularx}
\usepackage{graphicx}
\graphicspath{{./fig/}}
\usepackage{dcolumn}
\usepackage{pgfpages}
\usepackage{url}
\usepackage{comment}
\usepackage{lineno}
\usepackage[numbers,sort&compress]{natbib}

\usepackage{abstract}

\usepackage{setspace}
\usepackage[font=small, format=plain, labelfont=bf, up, textfont=normal, up, justification=justified,singlelinecheck=false]{caption}

\usepackage[affil-it]{authblk}

\let\OLDthebibliography\thebibliography
\renewcommand\thebibliography[1]{
\OLDthebibliography{#1}
\setlength{\parskip}{0pt}
\setlength{\itemsep}{0pt plus 0.3ex}
}

\title{Myopic Best Response as a Double-Edged Mechanism in Networked Social Dilemmas with Individual Solutions}
\author{Hirofumi Takesue\thanks{Electronic address: \texttt{hir.takesue@gmail.com}}}
\affil{Tokyo Metropolitan University}
\date{}

\begin{document}

\twocolumn[

\maketitle

\begin{onecolabstract}
Myopic best-response dynamics (MBRD) capture agents' bounded rationality and can generate evolutionary outcomes that differ from those produced by widely examined imitation dynamics. In this study, we apply MBRD to a three-strategy social dilemma---the snowdrift game with an individual solution---in which not only defection but also an individual solution that guarantees a safe, constant payoff can undermine cooperation. Monte Carlo simulations show that, on a square lattice, the evolutionary dynamics result in distinct equilibria, including the dominance of the individual solution, the coexistence of cooperators and defectors, or all-strategy coexistence. By combining simulations with a simple heuristic that approximates the transition condition between the dominance of the individual solution and the all-strategy coexistence, the analysis reveals a dual role of neighborhood size. Specifically, smaller neighborhoods can promote cooperation even when the individual solution is relatively inexpensive; however, achieving cooperation under these conditions requires greater benefits from cooperation. Notably, this hindrance to cooperation contrasts with evolutionary outcomes observed under imitation dynamics. Analysis of local strategy configurations explains the transition between the all-strategy coexistence and the coexistence of cooperators and defectors while showing that this transition is absent in a one-dimensional lattice. These observations indicate that the persistent availability of individual solutions constitutes an additional inhibiting factor of cooperation in populations of boundedly rational agents.
\\\\
\end{onecolabstract}
]
\saythanks

\section*{Introduction}
The evolution of cooperation has been studied across diverse disciplines, including the physical, biological, and social sciences. Evolutionary game theory provides a robust framework for analyzing the conflict between individual self-interest and collective welfare, revealing mechanisms that can explain the spontaneous emergence of cooperation \citep{Nowak2006, Roberts2021, Takacs2021, Xia2023, Rossetti2024, Han2026}. Recent studies have also explored how external institutions contribute to the maintenance of cooperation \citep{Chen2015c, Zhou2022, Shi2023, Yang2024}. Among the various factors influencing cooperation, networked interactions have been widely studied as a key mechanism facilitating its spontaneous evolution \citep{Nowak1992, Szabo2007, Roca2009, Perc2017}. Interactions with fixed neighbors within a network can sustain cooperation that would otherwise collapse in well-mixed populations.

The dynamics of strategy updating are a key component of evolutionary games on networks \citep{Roca2009}. Among these, imitation dynamics are the most extensively studied \citep{Szabo2007}. In imitation dynamics, agents compare their payoffs with those of their neighbors and are more likely to adopt a neighbor's strategy if it results in a higher payoff \citep{Szabo1998}. Such rules can serve as models of both biological evolution and social learning. Beyond imitation, researchers have examined how alternative updating dynamics affect cooperation in networked games. Notable examples include aspiration dynamics \citep{Chen2008}, moody conditional cooperation \citep{Grujic2012, Cimini2014}, reinforcement learning \citep{Ezaki2016}. Studies have also explored combinations of multiple updating rules. For instance, imitation dynamics have been combined with aspiration dynamics \citep{Xu2017a, Takesue2019a, Zhou2024}, conformity \citep{Szolnoki2015, Szolnoki2018a}, the death-birth process \citep{Szolnoki2018c}, reinforcement learning \citep{Lu2025a}, and logit dynamics \citep{Amaral2018, Danku2018, Li2021e}. Additionally, updating rules such as aspiration dynamics have been applied to the study of network evolution \citep{Huang2026}.

Among various updating dynamics, myopic best-response dynamics (MBRD), also known as the logit rule, has attracted research interest since early studies of evolutionary games on networks \citep{Roca2009, Roca2009a}. MBRD models agents who optimize payoffs using only current information about their neighbors' strategies, making it a useful starting point for studying bounded rationality. Importantly, MBRD produces patterns that differ significantly from those observed under imitation dynamics. Under MBRD, the improvement in cooperation seen with imitation dynamics occurs only in limited cases \citep{Roca2009b}. However, in the snowdrift game, MBRD in networked games can either facilitate or inhibit cooperation more than interactions in well-mixed populations, depending on parameter values, and cooperation can persist even in harsh conditions \citep{Sysi-Aho2005}. Additionally, in three-strategy games where imitation dynamics inhibit cooperation relative to well-mixed populations, MBRD can promote cooperation \citep{Lee2023}. MBRD also gives rise to characteristic spatial patterns. Unlike imitation dynamics, which maintain cooperation through the formation of large cooperative clusters through network reciprocity, MBRD supports the coexistence of cooperation and defection through checkerboard-like patterns on square lattices \citep{Szabo2010, Szabo2012, Shi2021a} and honeycomb-like patterns on triangular lattices \citep{Amaral2017}, particularly in the snowdrift game. These observations underscore the importance of considering MBRD when analyzing cooperation in networked evolutionary games.

In this study, we apply MBRD to the snowdrift game with an individual solution (SDGIS) \citep{Takesue2025}. The snowdrift game is a well-established model of social dilemmas \citep{Smith1982a}, in which individuals are incentivized to exploit their counterparts, but mutual free-riding leads to the worst outcome. This game has been applied to a broad range of contexts, including information security \citep{Hu2026}. Within the SDGIS framework, individuals may choose to pursue the required goal independently rather than select cooperation or defection in the dilemma game. This third option, referred to as the individual solution, introduces an additional barrier to cooperation, as individuals may prefer self-reliant actions rather than outright free-riding. The individual solution is a safe strategy because achieving the goal does not depend on others' cooperative efforts. However, such unilateral efforts are typically less efficient than outcomes achieved through cooperation (see Refs. \citep{Gross2019} and \citep{Takesue2026} for applications of the individual-solution concept). Owing to its safe yet inefficient nature, which resembles defection in stag hunt games, SDGIS can be viewed as a combination of snowdrift and stag hunt games. Evidence from human experiments further indicates that the availability of an individual solution often inhibits the emergence of cooperation \citep{Gross2019}. 

This study contributes to the literature on the evolution of cooperation by identifying the dual effects of networked interactions under MBRD. A previous simulation study showed that, when imitation dynamics are applied, networked interactions expand the parameter region in which cooperators can persist in the SDGIS \citep{Takesue2025}. This enhancement of cooperation arises primarily from three-strategy coexistence maintained through cyclic dominance, a mechanism also observed in voluntary social dilemmas \citep{Szabo2002, Szabo2002a, Hu2020, Szolnoki2020, Jia2024}. Conversely, our simulation results show that, under MBRD, networked interactions with a small neighborhood size often inhibit cooperation. We corroborate this observation using a simple heuristic that distinguishes among equilibrium states and provides threshold conditions for transitions from dominance of the individual solution to all-strategy coexistence. These findings provide further evidence of the distinctive effects of MBRD \citep{Sysi-Aho2005, Szabo2010, Szabo2012, Amaral2017, Shi2021a} and highlight the challenge posed by the individual solution as an additional barrier to cooperation among boundedly rational agents in networked games. 

\section*{Model}
We consider SDGIS, represented by the payoff matrix \citep{Takesue2025}:
\[
\begin{array}{c}
C\\ D\\ I\\
\end{array}
\begin{pmatrix}
b-c & b-2c & b-2c \\
b & 0 & 0 \\
b-c_I & b-c_I & b-c_I \\
\end{pmatrix}.
\]
The game consists of three strategies: cooperation ($C$), defection ($D$), and individual solution ($I$). Cooperation produces the benefits of a collective solution ($b$), which are enjoyed nonexclusively by \textit{both} interacting agents. The total production cost of the collective solution is $2c$. When both agents cooperate, this cost is shared equally, yielding a payoff of $b-c$. When only one agent cooperates, the cooperator bears the full production cost alone, resulting in the sucker's payoff of $b - 2c$. Agents who choose defection free ride on the benefit produced by a cooperating counterpart without incurring any cost and therefore obtain a payoff of $b$. However, when the counterpart does not cooperate, defectors receive a payoff of $0$. The condition $b - 2c > 0$ ensures that the interaction between $C$ and $D$ retains the defining structure of the snowdrift game. Selecting the individual solution yields a constant payoff of $b - c_I$, reflecting the low-risk nature of this strategy. Notably, the payoff from the individual solution is independent of the partner's action, because choosing this strategy guarantees achievement of the required goal in a manner comparable to mutual cooperation. Although the individual solution is less efficient than mutual cooperation, it is less costly than the unilateral provision of collective benefits. This relationship is captured by the assumption $c < c_I < 2c$. In the simulations presented below, $c$ is set to 1.

We explore evolutionary dynamics in which agents update their strategies by myopically selecting the strategy that maximizes their payoff. Agents are distributed across the vertices of a square lattice containing $N = L^2$ individuals, with periodic boundary conditions applied. At each elementary time step, one agent is randomly selected from the population. This agent tends to adopt the strategy $s$ that maximizes its payoff given the strategies of its four neighbors. Specifically, the probability of selecting strategy $s$ is expressed as follows:
\begin{equation*}
\frac{\exp(\beta U_s)}{\sum_{s' \in \{C, D, I\}} \exp(\beta U_{s'})}.
\end{equation*}
The probability of adopting strategy $s$ depends on the payoff $U_s$ that the strategy would yield, given the current strategies of the neighbors, as well as on the parameter $\beta$, which regulates the noisiness of strategy selection. This functional form of the Boltzmann distribution has been widely employed to model stochastic decision-making, including applications in policy selection in reinforcement learning \citep{Sutton1998} and the quantal response equilibrium in game theory \citep{Camerer2003}. In addition to payoff-based updates, strategies may also change through mutation, which occurs with probability $\mu$. In this case, the agent randomly adopts one of the three strategies, independent of the payoffs realized in local interactions.

We examined this evolutionary process using Monte Carlo simulations. At the start of each simulation run, a strategy is randomly assigned to each agent, and strategies subsequently evolve according to the dynamics described above. A Monte Carlo step (MCS) consists of $N$ elementary time steps, so that each agent has, on average, one opportunity to update their strategy during a single MCS. In typical simulations, the system undergoes a relaxation period of 20,000 MCSs to reach a steady state, after which data are collected over the subsequent 2,000 MCSs. The reported results represent the mean values obtained from four independent simulation runs. The frequencies of the three strategies are denoted as $f_C$, $f_D$, and $f_I$, corresponding to cooperation, defection, and the individual solution, respectively. The resulting equilibria are named according to the strategies that persist in the population; for example, a mixed-strategy equilibrium in which only cooperation and defection survive is called a $CD$-equilibrium.

\section*{Results}
\begin{figure}[tbp]
\centering
\vspace{5mm}
\includegraphics[width = 90mm, trim= 0 0 0 0]{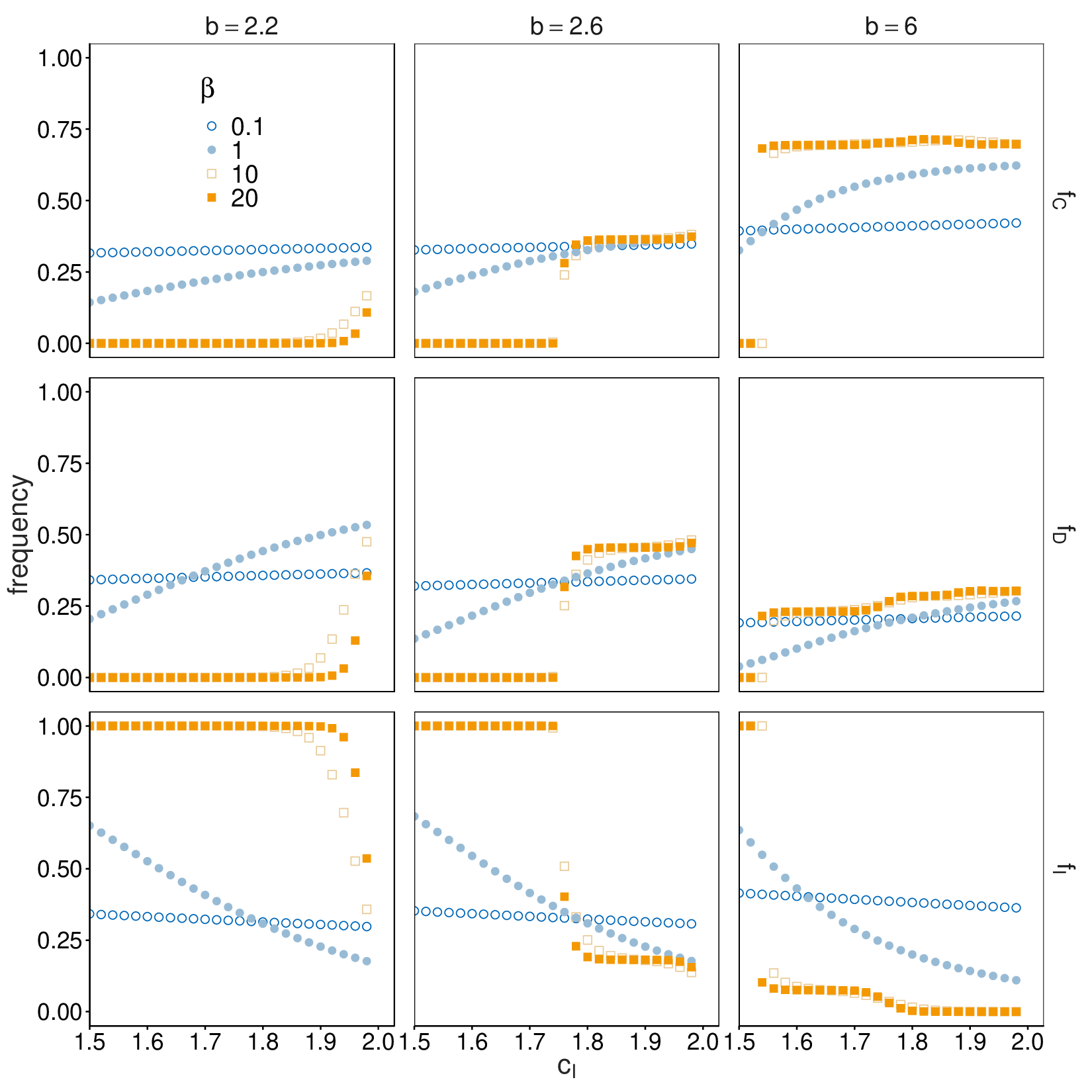}
\caption{\small Emergence of the $I$-, $CDI$-, and $CD$-equilibria under MBRD on a square lattice. As the cost of the individual solution $c_I$ increases, the $CDI$-equilibrium emerges, and this transition becomes more pronounced at higher selection intensities $\beta$. For larger values of the benefit parameter $b$, the same increase in $c_I$ instead leads to the $CD$-equilibrium. Simulation parameters: $L = 100$ and $\mu = 10^{-6}$.}
\label{cI_b_beta}
\end{figure}

The evolutionary dynamics give rise to three possible equilibria: the $I$-equilibrium, the $CDI$-equilibrium, or the $CD$-equilibrium. Figure~\ref{cI_b_beta} shows the resulting strategy frequencies as a function of the cost of the individual solution, $c_I$, for different values of the selection intensity $\beta$. As expected, the high cost of adopting the individual solution facilitates the invasion of cooperation and defection, leading to a transition from the $I$-equilibrium to the $CDI$-equilibrium. The distinction between the two equilibria becomes more pronounced at large selection intensities ($\beta = 10$ and $20$). The manner in which these transitions occur depends on the values of the benefit parameter $b$. When $b = 2.2$, cooperation and defection emerge gradually only when the value of $c_I$ approaches that of unilateral cooperation (i.e., $c_I = 2$), even under high selection intensity. Conversely, when $b = 2.6$, cooperation and defection appear at moderate values of $c_I$.

Simulations show that the $CD$-equilibrium emerges when the value of $b$ is large, such as $b = 6$. Unlike the cases with smaller values of $b$ (e.g., 2.2 and 2.6), the system exhibits the $I$-equilibrium, all-strategy coexistence, or the extinction of the individual solution as $c_I$ increases, with only a small fraction of agents adopting the individual solution due to mutations. The $CD$-equilibrium emerges when $\beta$ is sufficiently high.

\begin{figure}[tbp]
\centering
\vspace{5mm}
\includegraphics[width = 80mm, trim= 0 0 0 0]{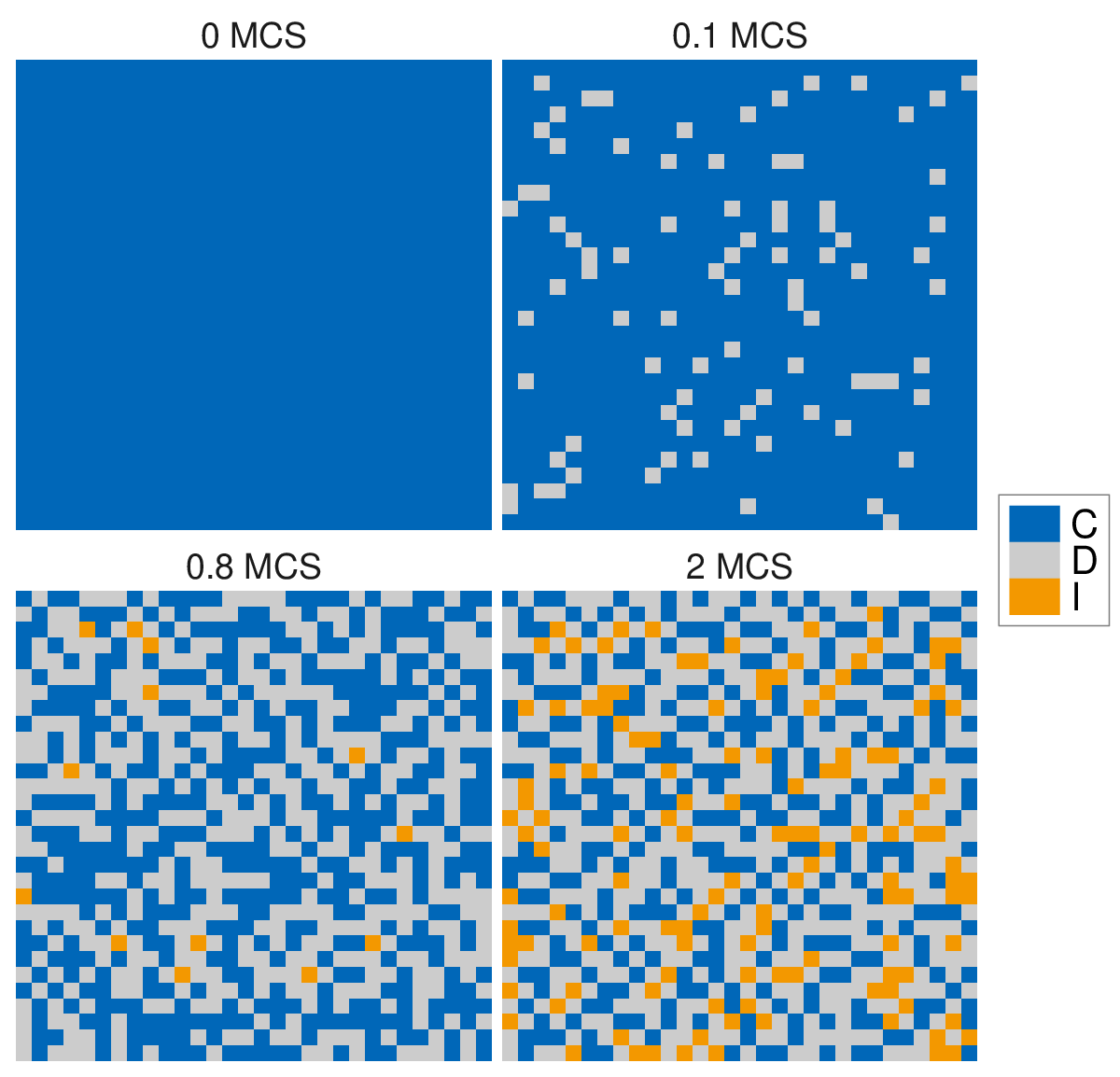}
\caption{\small Spatial coexistence of cooperation, defection, and individual solution on a square lattice. Defection yields the highest payoff when agents are surrounded by cooperators, whereas the individual solution yields the highest payoff when agents are surrounded by defectors. Simulation parameters: $L = 30, b = 2.6, c_I = 1.8, \beta = 10$, and $\mu = 10^{-6}$.}
\label{lattice_time}
\end{figure}

We first focus on the emergence of cooperation, specifically the transition between the $I$-equilibrium and the $CDI$-equilibrium. Since the $CDI$-equilibrium does not appear in well-mixed populations under replicator dynamics \citep{Takesue2025}, its emergence may require careful examination. Visualization of the dynamics on a square lattice elucidates the evolutionary process leading to the $CDI$-equilibrium (Figure~\ref{lattice_time}). To highlight local interactions, we consider a small system ($L = 30$) with all-$C$ initial states, which helps illustrate the evolutionary process. Agents surrounded by four cooperators adopt defection because it yields higher payoffs than cooperation by free-riding (0.1 MCS). As defectors accumulate, the individual solution begins to emerge, yielding higher payoffs than defection when surrounded by defectors (0.8 MCS). The individual solution dominates over cooperation in these contexts because it offers greater myopic payoffs in interactions with free-riders ($b - c_I > b - 2c$). These dynamics lead to the $CDI$-equilibrium observed in Figure~\ref{cI_b_beta}.

\begin{figure}[tbp]
\centering
\vspace{5mm}
\includegraphics[width = 80mm, trim= 0 0 0 0]{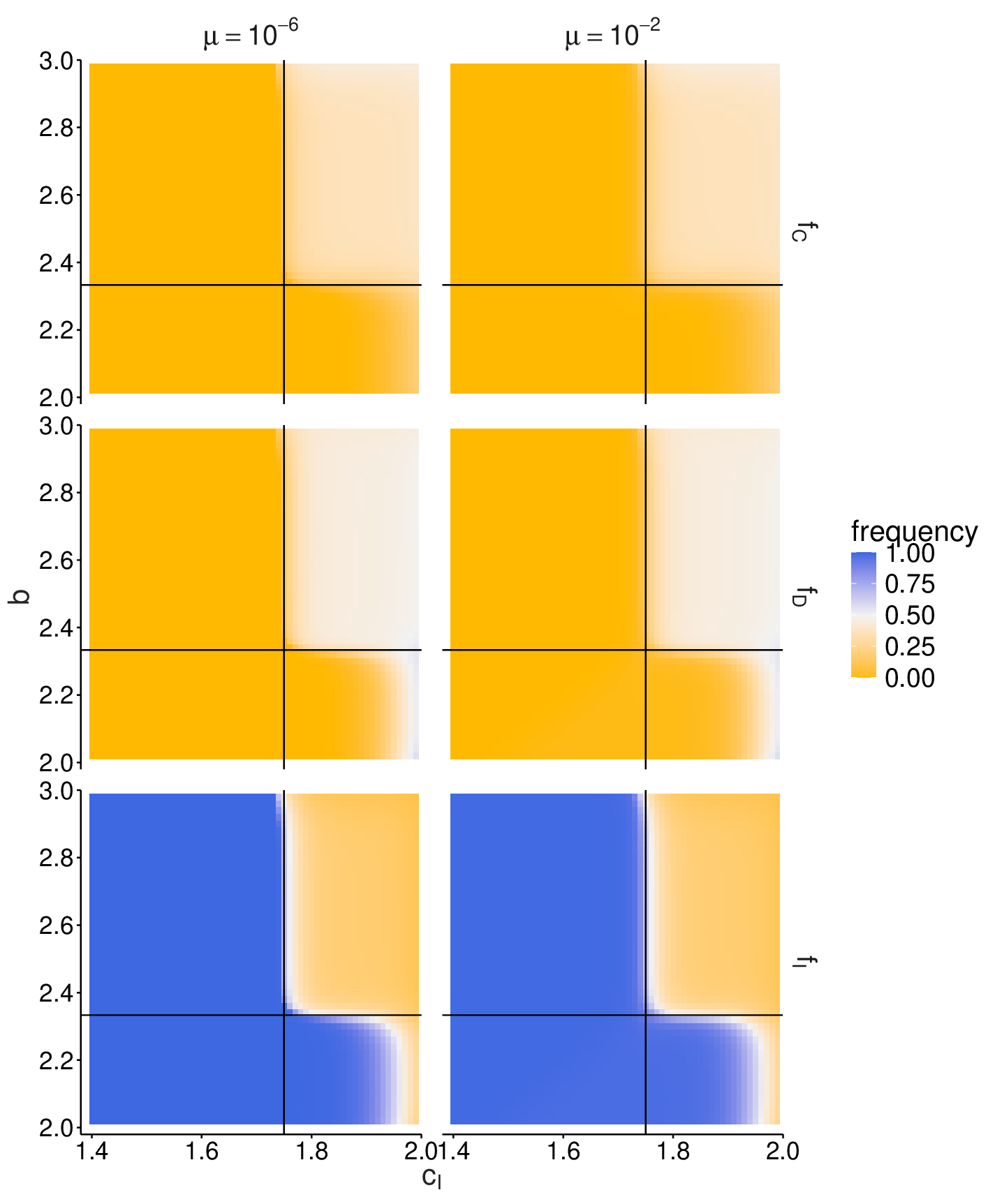}
\caption{\small Distinction between the $I$- and $CDI$-equilibria based on two threshold conditions. The heatmaps show that the $CDI$-equilibrium emerges when $c_I > 7/4$ and $b > 7/3$, with the black lines representing these two thresholds. For smaller values of $b$, cooperation and defection emerge only after the value of $c_I$ approaches that of unilateral cooperation (i.e., $c_I = 2$). Parameters: $L = 100, \beta = 10$, and $\mu = 10^{-6}$.}
\label{b_cI_phase}
\end{figure}

The emergence of the $I$- and $CDI$-equilibria is further illustrated using a heatmap. Figure~\ref{b_cI_phase} shows the frequencies of the three strategies as a function of the two payoff parameters: $b$ and $c_I$. As $c_I$ increases, the $CDI$-equilibrium emerges, as observed in Figure~\ref{cI_b_beta}. Additionally, large values of $b$ expand the parameter region in which the $CDI$-equilibrium occurs. More specifically, the conditions for the system to reach the $CDI$-equilibrium can be approximated by the two threshold lines shown in the figure: $c_I = 7/4$ and $b = 7/3$. When $b < 7/3$, the $CDI$-equilibrium emerges only after $c_I$ becomes sufficiently large for the individual solution to lose its relative advantage. Conversely, $c_I = 7/4$ sharply determines which of the two equilibria emerges when $b > 7/3$. Consequently, cooperation and defection persist in the upper-right region of the diagram, where both conditions are satisfied.

The two threshold conditions that distinguish the $I$-equilibrium from the $CDI$-equilibrium can be understood using a heuristic approach. In this study, we focus on the number of cooperators in a neighborhood ($n_C$) because, according to the payoff matrix, all strategies receive the same payoff in interactions with defection $D$ and the individual solution $I$. Table~\ref{pi_table} presents the payoff of each strategy categorized by $n_C$. The individual solution $I$ achieves the highest payoff when $n_C = 0$, indicating that a population absorbed into the $I$-equilibrium is unlikely to deviate from it. To prevent the system from settling into the $I$-equilibrium, cooperation must be selected by the neighbor of a cooperator when $n_C = 1$. Although defection may outperform the individual solution at $n_C = 1$, the neighboring cooperator would disappear unless cooperators also outperform the individual solution. Consequently, the $CDI$-equilibrium emerges when $4b - 7c > b$ and $4b - 7c > 4(b - c_I)$, which correspond to the conditions $b > 7c/3$ and $c_I > 7c/4$, consistent with the observations in Figure~\ref{b_cI_phase}. When cooperation outperforms the individual solution at $n_C = 1$, this also ensures the advantage of cooperators for $n_C \geq 2$, thereby permitting the emergence of defection as well. It should be noted that this heuristic is not strictly accurate, as the focal agent's strategy choice does not directly alter $n_C$, thus failing to depict the transitions between different $n_C$ configurations. Nevertheless, despite omitting the dynamics of local strategy frequencies, the heuristic provides a useful framework for understanding the transitions between the two equilibria.

\begin{table}[tbp] 
\centering
\caption{Payoff of each strategy conditional on the number of cooperators in the neighborhood ($n_C$)} \begin{tabular}{l>{\centering\arraybackslash}p{1.5cm}>{\centering\arraybackslash}p{1.5cm}>{\centering\arraybackslash}p{1.5cm}} 
\hline
 & $C$ & $D$ & $I$ \\ 
\hline 
$n_C = 0$ & $4b - 8c$ & $0$ & $4(b - c_I)$ \\ 
$n_C = 1$ & $4b - 7c$ & $b$ & $4(b - c_I)$ \\ 
$n_C = 2$ & $4b - 6c$ & $2b$ & $4(b - c_I)$ \\ 
$n_C = 3$ & $4b - 5c$ & $3b$ & $4(b - c_I)$ \\ 
$n_C = 4$ & $4b - 4c$ & $4b$ & $4(b - c_I)$ \\ 
\hline 
\end{tabular} 
\label{pi_table} 
\end{table}

\begin{figure}[tbp]
\centering
\vspace{5mm}
\includegraphics[width = 90mm, trim= 0 0 0 0]{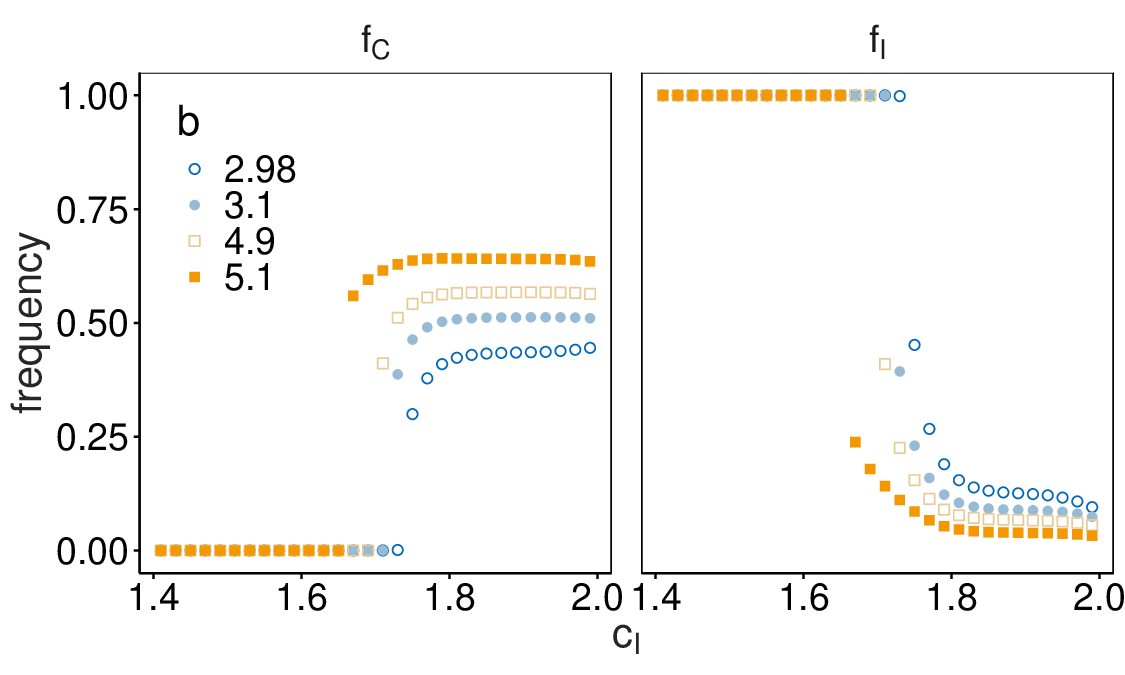}
\caption{\small Emergence of the $CDI$-equilibrium at values of $c_I$ smaller than those predicted by the heuristic when $b$ is large. Specifically, for $b > 3$, cooperation---and consequently the $CDI$-equilibrium---emerges at $c_I$ values below 7/4, which is the critical $c_I$ suggested by the heuristic. Simulation parameters: $L = 100, \beta = 10$, and $\mu = 10^{-6}$.} 
\label{b_cI}
\end{figure}

The conditions predicted by the simple heuristic do not apply when $b > 3$. Under this condition, the inequality $4b - 6c > 2b$ is satisfied (Table~\ref{pi_table}), indicating that cooperation yields higher payoffs than defection when $n_C = 2$. Consequently, the configuration with $n_C = 1$ becomes less important for predicting the survival of cooperation and defection. In this parameter regime, cooperation can persist even at smaller values of $c_I$, although its survival is not guaranteed. The outcomes illustrated in Figure~\ref{b_cI} reflect this behavior. Unlike the case of $b < 3$, whether $c_I$ exceeds 7/4 no longer reliably determines the emerging strategy composition.

We examine how different network structures influence evolutionary dynamics (see Ref. \citep{Flores2022} for a comprehensive analysis under imitation dynamics). In the analyses so far, we have focused on the von Neumann neighborhood on a square lattice, corresponding to a neighborhood size of $k = 4$. To examine the effects of limited neighbor interactions, we also examined one-dimensional lattices ($k = 2$), square lattices with the Moore neighborhood ($k = 8$), and well-mixed populations ($k = N - 1$). Using the heuristic based on the number of cooperators in the neighborhood (Table~\ref{pi_table}), the critical values of $b$ and $c_I$ can be derived similarly. Focusing on the case $n_C = 1$, a cooperator can emerge when $kb - (2k - 1)c > b$ and $kb - (2k - 1)c > k(b - c_I)$, which can be rewritten as $b > (2k -1)c/(k - 1)$ and $c_I > (2k - 1)c/k$. For the emergence of the $CDI$-equilibrium, the initial inequality implies that the produced benefit ($b$) need not be large when $k$ is large, whereas the second indicates that the cost of the competitor strategy ($c_I$) must increase with larger $k$. These results show that limited interactions arising from a small number of neighbors can both facilitate and hinder cooperation.

\begin{figure}[tbp]
\centering
\vspace{5mm}
\includegraphics[width = 80mm, trim= 0 0 0 0]{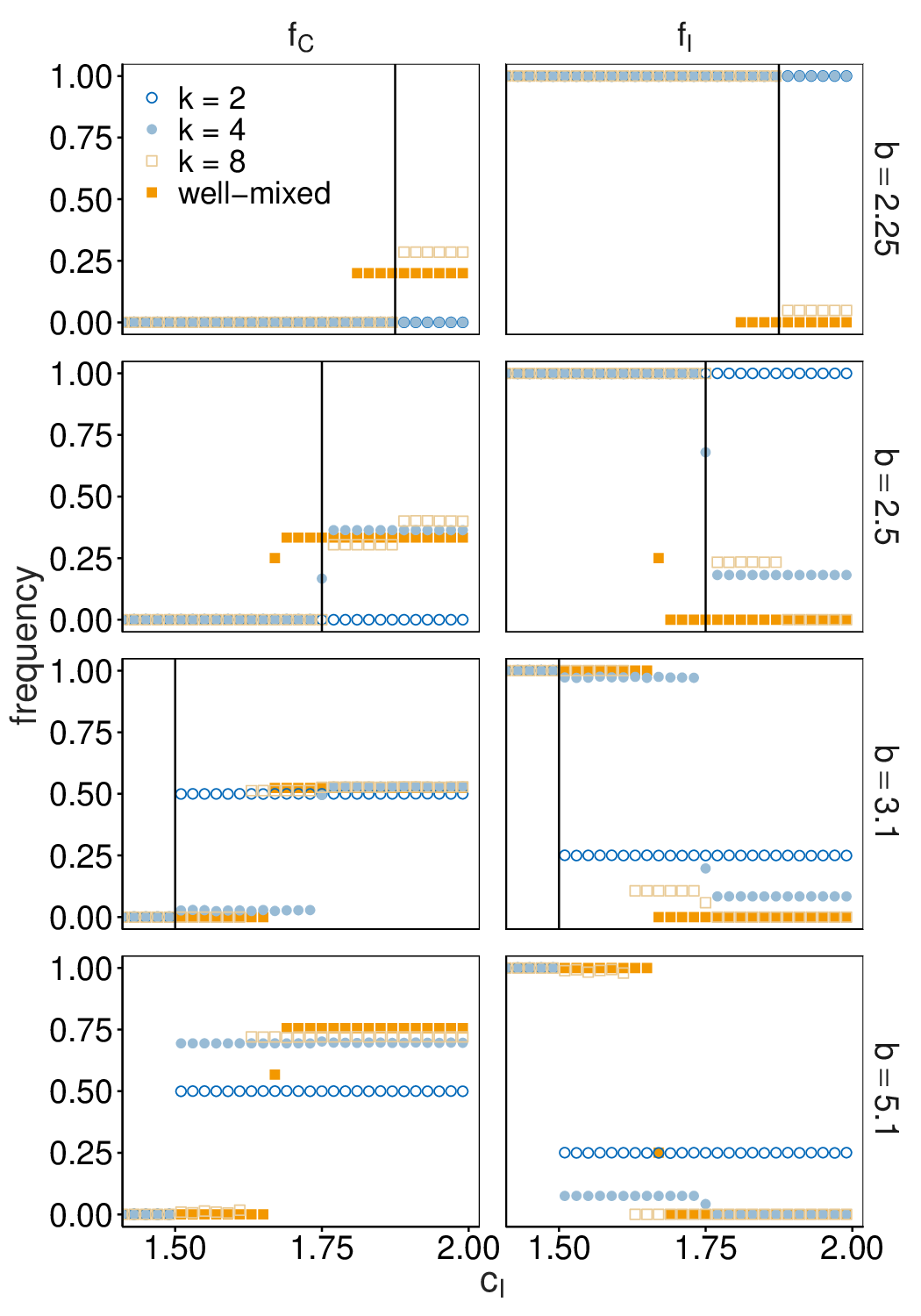}
\caption{\small Dual effects of limited interactions with a small number of neighbors on the emergence of cooperation (and the $CDI$-equilibrium). When the neighborhood size ($k$) is small, cooperation can emerge at relatively low values of the cost parameter $c_I$, but this emergence requires a large benefit $b$. The three lines indicate the critical values of $c_I$ that distinguish the $I$- and $CDI$-equilibria for $k = 8$, $4$, and $2$, ordered from the top panel to the bottom. The heuristic predictions are largely consistent with the simulation results, except in the case of a well-mixed population. This figure presents the results obtained under deterministic MBRD, implemented by assuming $\beta \to \infty$. Simulation parameters: $N = 10^4$ and $\mu = 10^{-6}$.} 
\label{cI_b_k_betaInf}
\end{figure}

The applicability of the heuristic across different neighborhood sizes is partially confirmed in Figure~\ref{cI_b_k_betaInf}. The figure illustrates the case of $\beta \to \infty$, which allows a precise evaluation of the performance of the heuristic, as agents adopt the strategy that maximizes their payoff with probability 1, given their neighbors' strategies. The lines in the figure indicate critical values of $c_I$ that separate the $I$- and $CDI$-equilibria. For the Moore neighborhood $k = 8$, the heuristic predicts that cooperation emerges when $b > 15/7$ and $c_I > 15/8$. In simulations with $b = 2.25$, a stepwise increase in $f_C$ and a corresponding decrease in $f_I$ are observed near the critical $c_I$ when $k = 8$. For $b = 2.5$, the jump in strategy frequencies is observed for $k = 4$, as previously observed in Figure~\ref{b_cI_phase}, where the value of $b$ exceeds the threshold for the von Neumann neighborhood (7/3), and cooperation emerges when $c_I$ reaches critical values of 7/4. A similar transition is observed for $k = 2$, where $b$ exceeds the required threshold of 3, and the transition is observed at the corresponding critical $c_I = 3/2$. Although the figure illustrates agreement with the heuristic regarding the critical $c_I$, additional simulations (not shown) confirmed that the heuristic also accurately predicts the critical values of $b$. 

The heuristic depends solely on whether a node has at least one cooperative neighbor; therefore, it is expected to operate similarly across different network structures. To assess its robustness, we conducted additional simulations using the network generation algorithm proposed by Watts and Strogatz \citep{Watts1998}, although these results are not shown. Specifically, we examined cases with rewiring probabilities of 0, 0.075, and 0.15. Across all cases, and for neighborhood sizes of $k = 2$, 4, and 8, the equilibrium transitions occurred at the same critical values of $b$ and $c_I$. 

In contrast to the outcomes observed in networked interactions, the results for the well-mixed population do not align with the predictions by the heuristic. When the heuristic is applied mechanically to the well-mixed case (with large $N$), it predicts the emergence of cooperation at $b > 2$ and $c_I > 2$. However, the figure shows that cooperation does not follow this predicted pattern. Instead, cooperation emerges across all panels when $c_I \leq 2$. This discrepancy indicates that a heuristic based on local strategy configurations does not apply to well-mixed populations.

\begin{figure}[tbp]
\centering
\vspace{5mm}
\includegraphics[width = 80mm, trim= 0 0 0 0]{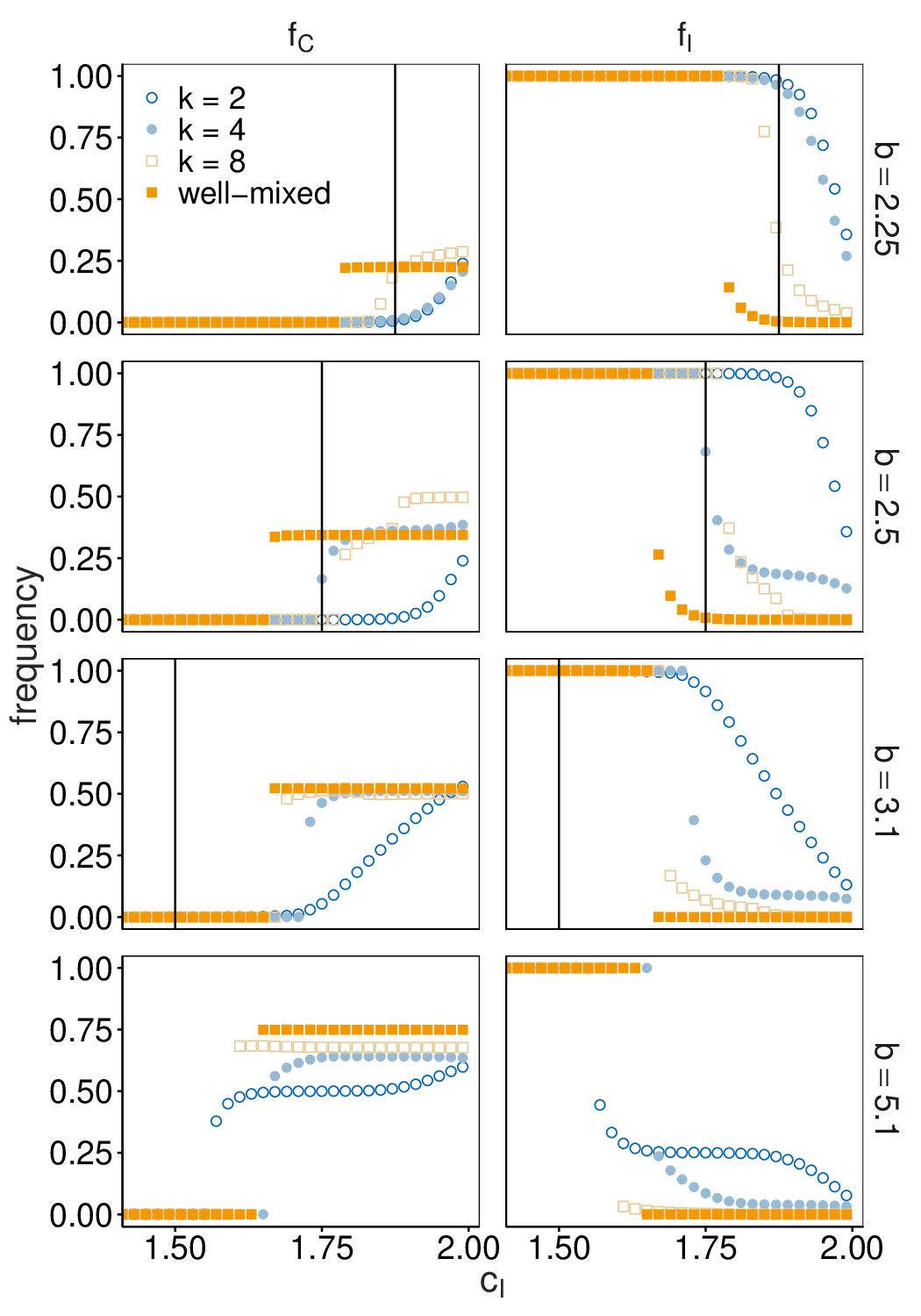}
\caption{\small Dual effects of small neighborhood sizes on the emergence of cooperation (and the $CDI$-equilibrium) under finite selection intensity ($\beta k = 40$). The patterns observed in Figure~\ref{cI_b_k_betaInf} are replicated. The three lines indicate the same information as that shown in Figure~\ref{cI_b_k_betaInf}. Simulation parameters: $N = 10^4$ and $\mu = 10^{-6}$.} 
\label{cI_b_k}
\end{figure}

A similar analysis was also conducted with $\beta < \infty$ (Figure~\ref{cI_b_k}). When $b = 2.25$ ($b = 2.5$), a similar transition from the $I$- to the $CDI$-equilibria is observed for $k = 8$ ($k = 4$). In these cases, cooperation emerges at similar values of $c_I$, although the transition occurs more gradually. A qualitative difference is observed for $k = 2$ when $b = 3.1$. Although the heuristic predicts that the benefits of cooperative behavior are sufficiently large under these conditions, cooperation emerges only gradually as $c_I$ increases. These observations indicate that finite values of $\beta$ can lead to deviations from the heuristic predictions in some cases.

The results in Figures~\ref{cI_b_k_betaInf} and \ref{cI_b_k} highlight the dual effects of interactions with a limited number of neighbors. Cooperation emerges at lower values of $b$ for $k = 8$ and in well-mixed populations. Additionally, $f_C$ often reaches higher values when the neighborhood size is large, as observed with large $b$ values in both figures. These observations indicate that limited local interactions can hinder cooperation. Conversely, cooperation can still emerge when the cost of adopting the competing individual solution is small. Specifically, local interactions facilitate cooperation for low $c_I$ once $b$ exceeds the critical value. It is important to note, however, that at finite selection intensity, the emergence of cooperation is delayed for small neighborhood sizes, which diminishes the merit of local interactions (e.g., $k = 2$ and $b = 3.1$ in Figure~\ref{cI_b_k}).

\begin{figure}[tbp]
\centering
\vspace{5mm}
\includegraphics[width = 90mm, trim= 0 0 0 0]{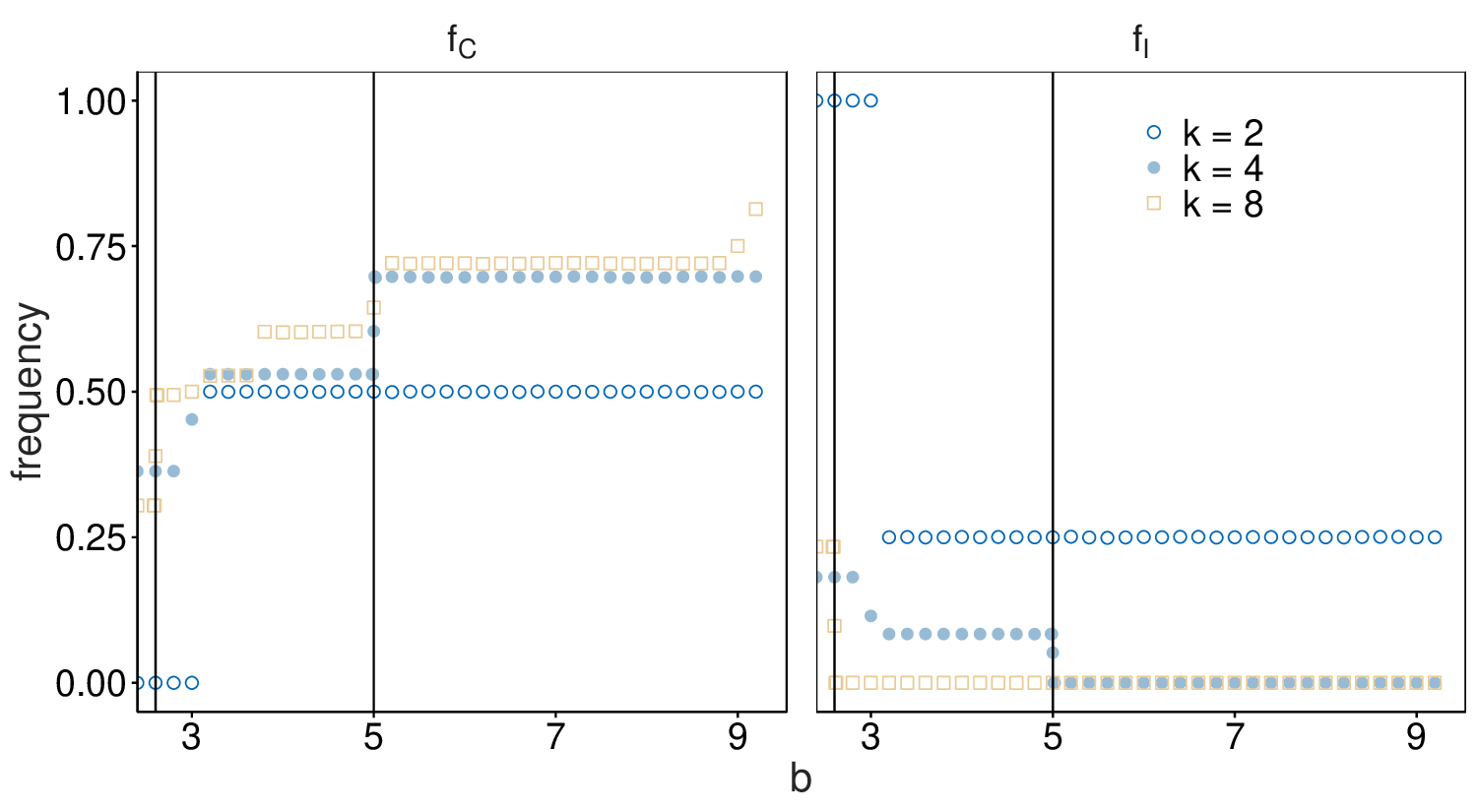}
\caption{\small Emergence of the $CD$-equilibrium and near disappearance of the individual solution in a square lattice, contrasted with its persistence in a ring lattice. The transition to the $CD$-equilibrium occurs at $b = 5$ for $k = 4$ and $b = 2.6$ for $k = 8$, whereas $f_I$ remains approximately 0.25 when $k = 2$. Deterministic strategy updating was applied. Simulation parameters: $N = 10^4, c_I = 1.85$, and $\mu = 10^{-6}$.} 
\label{b_k_bInf}
\end{figure}

The next analysis examines the transition from the $CDI$-equilibrium to the $CD$-equilibrium observed at large values of $b$ in Figure~\ref{cI_b_beta}. Figure~\ref{b_k_bInf} reports the strategy frequencies as a function of $b$ under deterministic updating ($\beta \to \infty$). The $CD$-equilibrium emerges at $b = 5$ for $k = 4$ and at $b = 13/5$ for $k = 8$. The frequency of the individual solution, shown in the right panel, becomes negligible once $b$ is sufficiently large. Conversely, a ring lattice with $k = 2$ does not exhibit this transition. Although the $CDI$-equilibrium emerges at $b \simeq 3$, as predicted by the heuristic, the individual solution continues to maintain a positive frequency, even at large values of $b$. 

\begin{figure*}[tbp]
\centering
\vspace{5mm}
\includegraphics[width = 110mm, trim= 0 0 0 0]{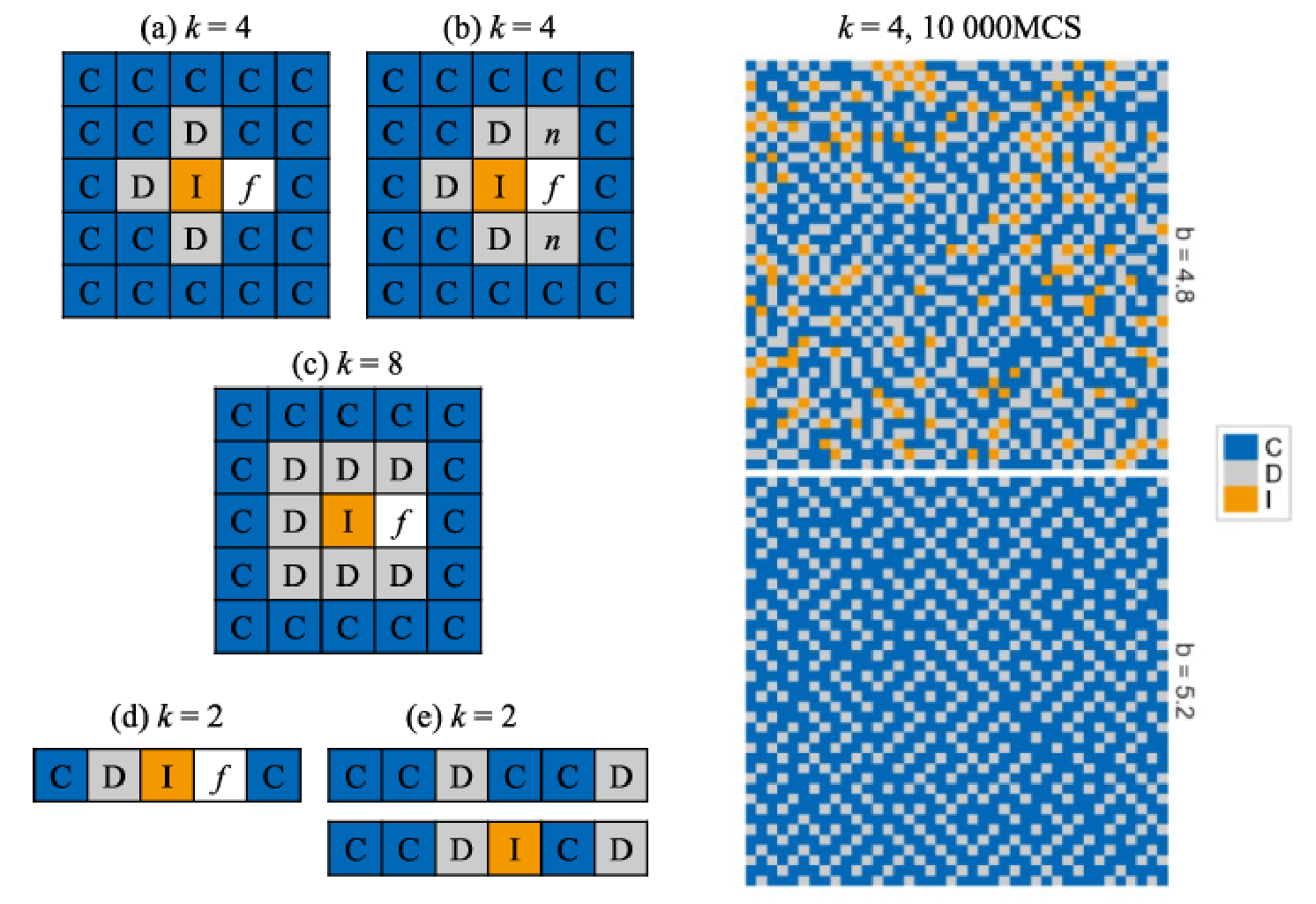}
\caption{\small Schematic representation of the transition between the $CDI$-equilibrium and the $CD$-equilibrium. The stability of the $CDI$-equilibrium depends on whether the neighbors surrounding $I$ continue to adopt $D$, as illustrated in cases (a) and (c); however, this configuration can become unstable over time, as shown in panel (b). This reasoning does not apply to a ring lattice (case (d)) due to the increased vulnerability of the $CD$-equilibrium to mutations (case (e)). The right panels show a 40 $\times$ 40 section of a simulation performed on a 200 $\times$ 200 lattice, where the transition to the $CD$-equilibrium produces a characteristic spatial pattern. Deterministic strategy updating was adopted. Simulation parameters: $c_I = 1.85$ and $\mu = 10^{-6}$.} 
\label{config}
\end{figure*}

Inspection of the local strategy configuration provides insights into the presence or absence of the transition. Figure~\ref{config} shows a schematic representation of the strategy configuration on a lattice. In a square lattice with the von Neumann neighborhood, $I$ can survive only when a node is surrounded by four $D$s. The advantage of $C$ over $I$ is reinforced as the number of cooperators increases ($n_C \geq 2$), provided that it is already advantageous when $n_C = 1$ (Table~\ref{pi_table}). Consequently, the environment that allows $I$ to emerge is restricted to configurations with $n_C = 0$. Case (a) in the figure illustrates this scenario, focusing on agent $f$, who is surrounded by three $C$s and one $I$. Agent $f$ must choose $D$ rather than $C$ to maintain the presence of $I$ because the agent currently using $I$ will switch to $C$ as soon as a cooperative neighbor emerges. Therefore, the $CD$-equilibrium emerges when $C$ outperforms $D$ in the case of $n_C = 3$, that is, when $4b - 5c > 3b$, which simplifies to $b > 5c$ (Table~\ref{pi_table}). This observation explains the transition observed at $b = 5$ in Figure~\ref{b_k_bInf}. A limitation of this reasoning is that it ignores dynamic effects. For instance, as presented in case (b), two neighbors of agent $f$, denoted as $n$, may adopt $D$, compelling agent $f$ to select $C$. This, in turn, can induce the central agent, originally adopting $I$, to switch to $C$. The right panels of the figure depict the spatial strategy configurations. In the $CDI$-equilibrium observed at $b = 4.8$, the individual solution is maintained primarily through protection by four free-riders. Conversely, the $CD$-equilibrium observed at $b = 5.2$ shows a structured pattern characterized by free-riders being surrounded by cooperators, whereas cooperators have both cooperative and defective neighbors.

The transition in the Moore neighborhood can be interpreted in a similar manner. In case (c), the central agent is surrounded by eight free-riders. This cluster remains stable against invasion when agent $f$, with three cooperative neighbors, continues to select $D$. (The center agent's diagonal neighbors are more likely to select $D$ because of the large number of cooperative neighbors.) The case of $n_C = 3$---that is, $3(b - c) + 5(b - 2c) > 3b$---corresponds to the critical value $b = 13/5$ in Figure~\ref{b_k_bInf}. 

Lastly, we focus on the ring lattice illustrated in case (d). In contrast to the von Neumann and Moore neighborhoods, the simple heuristic analysis does not predict the outcomes of this structure. Applying the same logic suggests that the $CD$-equilibrium should emerge when $(b - c) + (b - 2c) > b$. However, simulations indicate that the individual solution persists in the population. The underlying mechanism of this discrepancy arises from the instability of adjacent cooperators. Case (e) shows the $CD$-equilibrium in the ring lattice, which emerges when cooperators outperform free-riders in the case of $n_C = 1$, that is, $(b - c) + (b - 2c) > b$. In this case, cooperators positioned between $C$ and $D$, and free-riders located between two $C$s, achieve maximal payoffs. However, this equilibrium is vulnerable to noise. If a single cooperator is replaced by an agent adopting $I$, the adjacent cooperator can adopt $I$ because $n_C = 0$. This vulnerability of adjacent cooperators explains the persistence of the $CDI$-equilibrium in the ring lattice. 

\section*{Discussion}
This study examined the evolution of cooperation in SDGIS on networks under MBRD. The SDGIS is the snowdrift game with the third strategy, the individual solution, which provides a constant payoff. Monte Carlo simulations revealed the emergence of three distinct equilibria—the $I$-equilibrium, the $CDI$-equilibrium, and the $CD$-equilibrium, depending on agents' myopic selection of the best response to the current strategies of their neighbors. A heuristic that considers the local distribution of strategies explains the transition between the $I$-equilibrium and the $CDI$-equilibrium, particularly when the benefits of cooperative efforts are moderate. Simulations also showed that interactions with a limited number of neighbors---that is, a low network degree---have dual effects. Specifically, a low degree facilitates the emergence of cooperation when the cost of the individual solution is low. However, when the number of neighbors is small, the threshold for benefits required to sustain cooperation increases. The analysis of local strategy configurations also clarifies the mechanisms underlying transitions, or their absence, between the $CDI$-equilibrium and the $CD$-equilibrium. 

A comparison with previous studies highlights the crucial role of strategy updating dynamics. Parameter regions in which cooperation was observed expanded in SDGIS when network interactions and imitation dynamics were combined \citep{Takesue2025}. Conversely, MBRD does not support the emergence of cooperation, particularly when the benefits of cooperative efforts are small. This observation provides a novel example showing that, under MBRD, interactions on networks do not necessarily promote cooperation \citep{Roca2009b}. Furthermore, this study reinforces the claim that MBRD and imitation dynamics can have opposing effects on the emergence of cooperation \citep{Lee2023}. The introduction of the third strategy also generated novel spatial patterns. Unlike cooperation sustained by checkerboard-like patterns in the two-strategy snowdrift game \citep{Szabo2010, Szabo2012, Shi2021a}, free-riders are surrounded by cooperators, whereas cooperators maintain both cooperative and defective neighbors (Figure~\ref{config}). This type of configuration, previously observed in the snowdrift game on a triangular lattice \citep{Amaral2018}, is reproduced in SDGIS on a square lattice.

This study concludes by highlighting its limitations and identifying potential avenues for future research. This study focused on MBRD, but a more cognitively demanding strategy updating rule could be explored. Reinforcement learning is a promising approach that is increasingly attracting the attention of the research community \citep{Xie2026}. Unlike MBRD, which promotes cooperation through checkerboard-like spatial patterns, reinforcement learning often maintains cooperation through compact clusters of cooperators, similar to patterns observed under imitation dynamics \citep{Zhao2024, Kang2025}. This difference suggests that the spatial patterns of cooperation on networks can reveal valuable information about the underlying microscopic updating dynamics. Research that systematically examines the role of different strategy updating dynamics may therefore provide deeper insights into the mechanisms driving the evolution of cooperation.


\end{document}